\documentclass[twocolumn,english,prl,superscriptaddress]{revtex4}
\usepackage{amsmath,amssymb,amsfonts}
\usepackage{CJK}
\usepackage{bm}
\usepackage{tikz}
\usepackage{pgfplots}
\usetikzlibrary{matrix,fit}
\usepackage{graphicx} 
\usepackage{braket}
\usepackage{subfig}
\usepackage{caption}
\usepackage{mathrsfs}
\usepackage{float}
\usepackage[utf8]{inputenc}
\usepackage{pgflibraryarrows}
\usepackage{balance}
\usepackage{lipsum} 
\usepackage{pgflibrarysnakes}

\begin{document}
\renewcommand{\raggedright}{\leftskip=0pt \rightskip=0pt plus 0cm}
\captionsetup[figure]{name={FIG.},labelsep=period}
\title{Topological Disclination Pump}

\author{Bi-Ye Xie}\thanks{These authors contributed equally to this work.}
\affiliation{Department of Physics,
The University of Hong Kong, Pokfulam Road, Hong Kong, China}

\author{Oubo You}\thanks{These authors contributed equally to this work.}
\affiliation{Department of Physics,
The University of Hong Kong, Pokfulam Road, Hong Kong, China}

\author{Shuang Zhang}
\email[]{shuzhang@hku.hk}
\affiliation{Department of Physics, 
The University of Hong Kong, Pokfulam Road, Hong Kong, China}
\affiliation{Department of Electrical and Electronic Engineering, The University of Hong Kong, Pokfulam Road, Hong Kong, China}


\begin{abstract}
A topological pump enables robust transport of quantized particles when the system parameters are varied in a cyclic process. In previous studies, topological pump was achieved in homogeneous systems guaranteed by a topological invariant of the bulk band structure when time is included as an additional synthetic dimension. Recently, bulk-boundary correspondence has been generalized to the bulk-disclination correspondence, describing the emergence of topological bounded states in the crystallographic defects protected by the bulk topology. Here we show the topological pumping can happen between different disclination states with different chiralities in an inhomogeneous structure. Based on a generalized understanding of the charge pumping process, we explain the topological disclination pump by tracing the motion of Wannier centers in each unit cell. Besides, by constructing two disclination structures and introducing a symmetry-breaking perturbation, we achieve a topological pumping between different dislocation cores. Our result opens a route to study the topological pumping in inhomogeneous topological crystalline systems and provides a flexible platform for robust energy transport.

\end{abstract}
\maketitle

{\it{Introduction}}.--- The concept of topological pump was originally introduced by Thouless in 1983~\cite{Thouless} which describes a quantum phenomenon that a particle can be transported via a slow cyclic variation of an one-dimensional potential. A similar classical analog of topological pump is the well-known Archimedes screw~\cite{AS} that can pump water by turning a screw-shaped surface inside a pipe. Intriguingly, different from the Archimedes screw where the amount of water is continuously pumped in a slow turning process, the topological pump can only happen in a discrete manner, with a quantized number of particles pumped in each cycle, which is characterized by a topological quantum number~\cite{Thouless}. Charge pump~\cite{Niu1,Niu2,pump4,pump5} has been investigated in various systems such as quantum dots~\cite{pump1,pump2,pump3}, 1D acoustic channels~\cite{pump3} and quasicrystals~\cite{pump6}. Recently, topological pump~\cite{TP0,TP1} was realized in cold atom systems~\cite{TPE1,TPE2}, which triggered various experimental studies on topological pump including exploration of nonlinear effect~\cite{TPE3}, higher-dimensional quantum Hall physics~\cite{4D1,4D2}, and topological pumping of elastic waves~\cite{TPE4} and acoustic waves~\cite{acoustic1,acoustic2}. Recently, Benalcazar et al.~\cite{HTP} generalized the conventional pump to higher-order topological systems and observed transport of photons between corner states, boundary states of higher-order topological insulators. Nevertheless, all these studies were limited to homogeneous lattice structures and focused on pumping of exterior boundary states. 2D defects such as the dislocations and disclinations are diverse and very common in lattices, showing fascinating physical properties such as the existence of topologically protected bound states \cite{TD}. So far, the theory of topological pump has not been investigated in inhomogeneous systems with real-space defect structures.

In parallel, bulk-boundary correspondence, one of the most important notion of topological physics~\cite{TI1,TI2,TI3,TI4,TI5,TI6,TI7} that describes the connection between bulk topological invariants and the emergence of gapless boundary states, has been widely studied~\cite{BBC1,BBC2,BBC3}. Recently, the bulk-boundary correspondence has been generalized to the bulk-disclination correspondence~\cite{BDC1,BDC2,BDC3,BDC4}, i.e., a bound state can emerge at the disclination cores in the absence of any spectral signatures, which is the manifestation of symmetry-protected higher-order topological crystalline phases. Importantly, these bounded states are robust against perturbations that preserve the bulk topology. The bulk-disclination correspondence has been experimentally demonstrated in 2D metamaterials and photonic crystals~\cite{BDC3,BDC4}. However, as these observations are obtained in purely static systems, it remains unclear what the influence on the disclination states would be from a dynamic modulated bulk topology which is crucial for distinguishing the topological boundary states from the locally induced defect states.

\begin{figure*}
\centering
\includegraphics[scale=0.56]{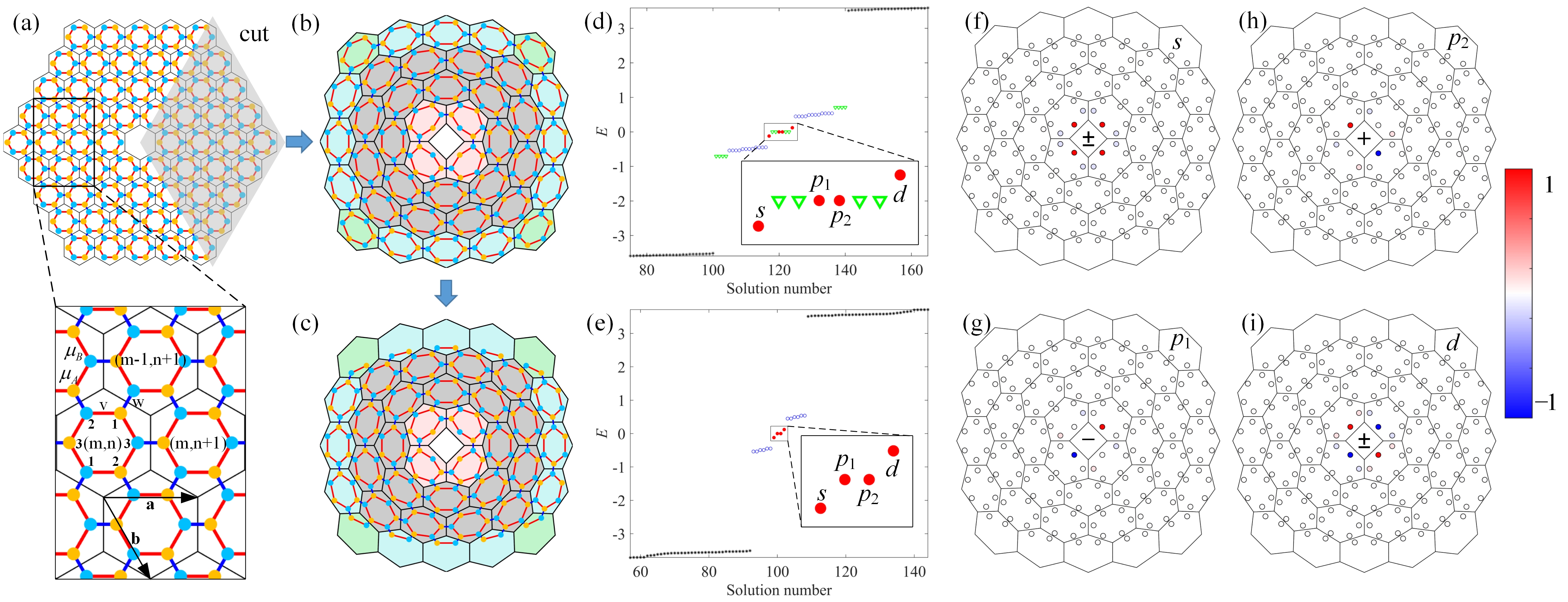}
\captionsetup{format=plain, justification=raggedright}
\caption{The formation of disclinations and eigenstates. (a) The original $C_6$ symmetric homogeneous lattice. The shadow area is removed to construct a $C_4$ symmetric disclination structure. The parameters in $\mathcal{H}$ is shown in the lower panel. The orange (blue) dots represent A (B) sublattices. The red (black) line represent the inter-cell (intra-cell) coupling. (b) A $C_4$ symmetric disclination structure. (c) Disclination structure with removing some boundary sites. (d) The eigenstates of (b) where there are bulk states (black stars), edge states (blue dots), corner states (green triangles) and disclination states (red dots) whose field distributions are mainly distributed in the grey, blue, green and red unit cells as shown in (b). (e) The eigenstates of the disclination structure with removing some boundary sites. There are no zero-energy corner states. (f)-(i) the field distributions for $s$, $p_1$, $p_2$, and $d$ states respectively. The symbols $+$ and $-$ mark the positive and negative chiralities.}
\label{fig:1}
\end{figure*} 

In this Letter, we show that topological pump can occur between different disclination states. We first consider a tight-binding realization of a distorted two-dimensional (2D) lattice model with $C_6$ symmetry for achieving a $C_4$ symmetric crystalline structure with chiral-symmtric disclinations. Then we analyse the higher-order bulk topology and show the presence of four disclination states as a consequence of the bulk-disclination correspondence. Importantly, we numerically demonstrate a topological pump between two local disclination states by presenting the spectral flow and field distributions with respect to varying nearest-neighbor coupling and onsite energies. Furthermore, we design a system with two disclinations and introduce perturbations that break the chiral symmetry inside each disclination to demonstrate a topological pump between spatially separated disclination states, a manifestation of the unique feature of the topological origin rather than local-structure-induced defect states. Our results extended the study of topological pump into inhomogeneous systems and show that the quantized topological transport is robust against by geometric distortions.

{\it{The disclination states and bulk-disclination correspondence}}.---We start from a 2D spinless tight-binding model with nearest-neighbour couplings as shown in Fig. 1(a). We neglect the interaction between electrons and thus this model describe a single particle system. The Hamiltonian of this model is 
\begin{equation}
\begin{aligned}
\mathcal{H}(t)&=v(t)\sum\limits_{m,n=1}^{N}\sum\limits_{i\neq j}(c^\dagger_{m,n,B,i}c_{m,n,A,j}+h.c.)\\
&+w(t)\sum\limits_{m,n=1}^{N-1}(c^\dagger_{m-1,n+1,B,1} c_{m,n,A,1}\\
&+c^\dagger_{m+1,n,B,2} c_{m,n,A,2}+c^\dagger_{m,n+1,B,3} c_{m,n,A,3}+h.c.)\\
&+\mu(t)\sum\limits_{m,n=1}^{N}(c^\dagger_{m,n,A}c_{m,n,A}-c^\dagger_{m,n,B}c_{m,n,B})
\end{aligned}
\end{equation}

Where $v(t)$, $w(t)$, and $\mu{(t)}$ are the inter-cell coupling, the intra-cell coupling, and the onsite potential respectively which are all real, positive functions.  $c^\dagger$ ($c$) is the creation (annihilation) operator for spinless particles and h.c. stands for complex-conjugate terms. $n$ and $m$ are integers labelling the sequence number of unit cells along the lattice vector $\bm{a}$ and $\bm{b}$ respectively. $i, j=1,2,3$ label different lattice sites within a unit cell. $A$ and $B$ represent two sublattices in the model. The evolution of the system can be captured by considering a parameter $t$, usually regarded as the time degree of freedom. This Hamiltonian describes a graphene-like lattice with alternating inter-cell and intra-cell coupling as well as staggered onsite potentials on different sublattices. We first study the topological physics of the Hamiltonian without onsite energy potential $\mathcal{H}_0=\mathcal{H}-\mu(t)\sum\limits_{m,n=1}^{N}(c^\dagger_{m,n,A}c_{m,n,A}-c^\dagger_{m,n,B}c_{m,n,B})$. $\mathcal{H}_0$ has been shown to have non-trivial (trivial) higher-order topological crystalline phases with helical edge states and spinful corner states (for an open boundary) if $v<w$ ($v>w$)~\cite{topo1}.

We next consider a finite size structure with $N=5$ ($N$ stands for the number of unit cells at one edge of the structure) as shown upper panel in Fig. 1(a). This structure has a global $C_6$ symmetry and can be formed by rotating one sector six times with respect to the center. To construct the disclination structure, we cut out two sectors and glue the boundaries while preserving the chiral (sublattice) symmetry to form a deformed structure (DS) with a global $C_4$ symmetry as shown in Fig. 1(b). This inhomogeneous DS inherits the topological properties of the original periodic lattice. To prove this, we calculate the eigenmodes of the DS (see Fig. 1(d) with detailed discussion on the band structure of the periodic lattice in Section I in SI~\cite{SUP}) and show that for $v<w$ (specifically, we set $v=0.5$ and $w=2$.), there is an energy gap between bulk states (dark points in Fig. 1(d)) and within the gap, there are topological boundary states such as edge states (blue circles in Fig. 1(d)) and corner states (green triangles in Fig. 1(d)). Moreover, we find four in-gap states localized at the disclination cores (red dots in Fig. 1(d)) which are disclination states. Disclination states have been previously studied for their fractional charges in DS with $C_3$ and $C_5$ global symmetries. Unfortunately, all these DSs have no chiral symmetry since one cannot define A and B sublattice globally in these structures with an odd number of sectors being cut out. On the other hand, the chiral symmetry is preserved in our DS.

The topological phases of $\mathcal{H}_0$ in this $C_6$ symmetric lattice is characterized by the topological invariant $\chi^{(6)}=([M],[K])$ with  $[M]=\#M_1-\#\Gamma^{(2)}_1$ and $[K]=\#K_1-\#\Gamma^{(3)}_1$~\cite{topo2}. Here $\#M_1$ and $\#\Gamma^{(2)}_1$ ($\#K_1$ and $\#\Gamma^{(3)}$) are the number of states below the bandgap with $C_2$ eigenvalue $M_1=+1$ and $\Gamma^{(2)}_1=+1$ at the M ($\Gamma$) point  (with $C_3$ eigenvalue $K_1=+1$ and $\Gamma^{(3)}_1=+1$ at the K ($\Gamma$) point) of the Brillouin zone respectively. For $v<W$ ($v>w$), we have $\chi^{(6)}=(2,0)$ ($\chi^{(6)}=(0,0)$), which indicates a topologically non-trivial (trivial) phase with Wannier centers (WCs) located at the edges (centers) of unit cells~\cite{HTP}. The non-trivial topological phase is an obstructed atomic insulating phase. Disclination states in the DS inherit the above topology with only a small drift of the WCs from the high-symmetry point (Wyckoff positions) of unit cells due to the slightly broken $C_6$ symmetry of unit cells in the DS~\cite{topo3}. Subsequently, one can define a second topological index to characterize the charge of disclination states as $Q=\frac{\Omega}{2\pi}(\frac{3}{2}\chi_M-\chi_K) module 1$ where $\Omega$ is the Frank angle characterizing the disclinations~\cite{topo3}. For our case, $\Omega=\frac{2\pi}{3}$ and thus $Q=0$ which means that there is no fractional charges. We here emphasize that disclination states are not necessarily accompanied by fractional charges. Instead, disclination states are ensured by the bulk-disclination correspondence. Nevertheless, the slight broken $C_6$ symmetry of unit cells will always induce a small non-integer charge of the disclination states.

The four disclination states can be labelled as $s$, $p_1$, $p_2$, and $d$ states based on their spatial distributions shown in Figs. 1 (d)-(i). Among these four states, only $p_1$ and $p_2$ are the eigenstates of the chiral symmetry operator with opposite chiralities , hence they form chiral symmetric partners to each other (see Figs. 1(h)-(g)). Without loss of generality, we define the chiralities of $p_1$ and $p_2$ as $+1$ and $-1$ respectively (see Figs. 1(h)-(g)). As a consequence, $p_1$ and $p_2$ are pinned at zero-energy modes which is protected by chiral symmetry while $s$ and $d$ are shifted from the zero energy (see Figs. 1(d)-(e)). We note that there are also four outside zero-energy corner states (represented by green triangles in  Fig. 1(d)) (see detailed discussions of the field distributions of corner states in Section II in SI~\cite{SUP}). Nevertheless, we can eliminate these states by changing the side geometry as shown in Fig. 1(c). After removing several lattice sites at the outsider boundary, all zero-energy corner states disappear and only two $p$-type disclination states remains (see Fig. 1(c)).

\begin{figure}
\centering
\includegraphics[scale=1]{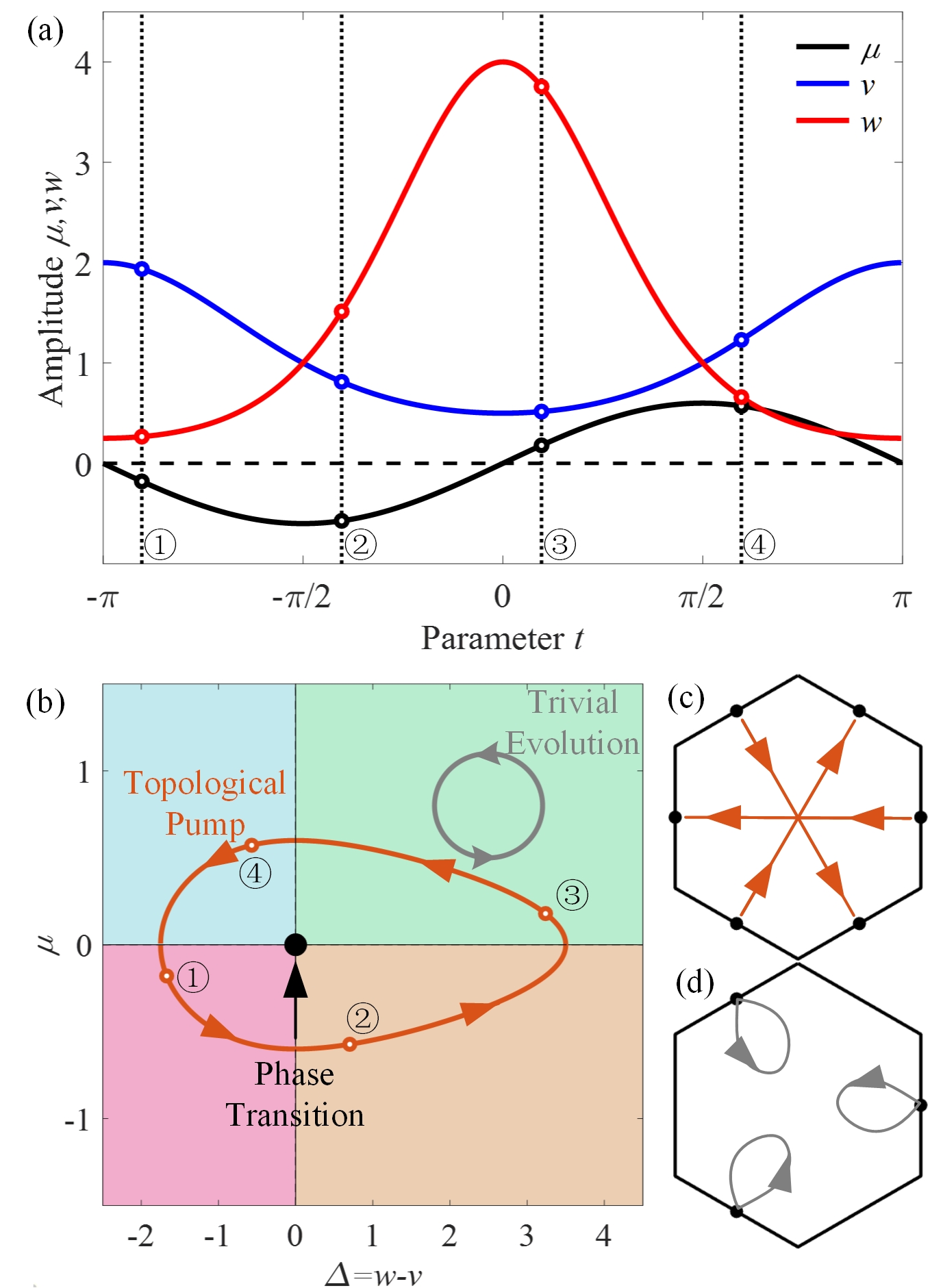}
\captionsetup{format=plain, justification=raggedright}
\caption{Parameters evolutions for the topological pump. (a) The evolution of $\mu$ (black line), $v$ (blue line), and $w$ (red line) as functions of $t$ in a period. (b) Schematic of the evolution of the Hamiltonian $\mathcal{H}$ in the parameter space. For topological pump, the trajectory encircle the gapless phase transition point (orange line) while for trivial evolution, the trajectory do not. (c) The motion of three Wannier centers (black dots) during the pumping process (yellow lines). (c) The motion of three Wannier centers (black dots) during the trivial evolution process (grey lines).}
\label{fig:1}
\end{figure}

{\it{Local topological pump between disclination states}}.---In this section, we discuss the topological pumping process between two disclination states localized at the same disclination core. We consider an adiabatic process in which when parameter $t$ continuously varies from $-\pi$ to $\pi$, $v(t)$, $w(t)$, and $\mu{(t)}$ vary in a way that does not close the bandgap as shown in Fig. 2(a). Specifically, we have

\begin{equation}
\begin{aligned}
v(t)=0.5^{\mathrm{cos}(t)}, w(t)&=4^{\mathrm{cos}(t)}, \mu(t)&=0.6\mathrm{sin}(t)
\end{aligned}
\end{equation}
For $t\in(-\pi, \pi)$. As a consequence, the evolution of the Hamiltonian $\mathcal{H}$ forms a closed loop (the orange circle in Fig. 2(b)) encircling the gapless point in the parameter space (see Fig. 2(b)) and thus induces a topological pump. We here emphasize that the selection of the evolution functions of parameters are not unique and the pump occurs as long as the evolution path enclose the gapless point. The non-zero values at two axes in the parameter space represent two orthogonal mass terms in the effective Hamiltonian around the gapless phase transition point.  We note that a closed loop (the grey circle in Fig. 2(b)) in the parameter space that does not circle the gapless point is a trivial evolution that will not contribute to the transportation of charges since this loop can be continuously deformed into a point. 

The topological pump can be intuitively understood by tracing the motion of WCs within a unit cell in the following way~\cite{HTP}(see Fig. 2(c) and 2(d)): For the initial configuration of the evolution with $v<w$, WCs are located at edges of unit cells which corresponds to the topologically non-trivial phase. As the evolution starts, WCs travel towards the center of the unit cell. When $v>w$, WCs locate at centers of unit cells and the system is in topologically trivial phase. Finally, at the end of the evolution, $v<w$ and WCs move to the opposite side of the edge. Since the position of a Wannier center is the expected value of the position operator under the Wannier function basis, it characterizes the position of the charge center. Therefore, the above period realizes a quantized topological transport of charges between different disclination core sites in real space. For the trivial parameter evolution, WCs circle back to their original positions, and hence there is no transport of charges.

One of the most important signatures of a topological pump is the spectral flow of boundary states~\cite{TPE1,TPE2} (or the disclination states here) in the energy spectrum. The evolution of the eigenstates of the DS shown in Fig. 1(c) with respect to $t$ is presented in Fig. 3(a) which shows the evolution spectrum of two chiral symmetry partners linearly intersecting each other around the degenerate point $t=0$, clearly demonstrating a spectral flow of the disclination states. We further prove this by presenting local field distributions of wavefunctions as shown in Fig. 3(b)-(d). An initial zero-energy disclination state localized at sublattice A with chirality $+1$ is prepared at the beginning of the pumping process. As $t$ increases, the disclination state is gradually pushed into the bulk spectrum (Fig. 3(c)). Near the end of the pumping process, the disclination state again emerges from the bulk spectrum with an opposite chirality $-1$ localized at the $B$ sublattice (Fig. 3(d)). The pumping process transport energy between two sublattices with the same disclination core.

\begin{figure}
\centering
\includegraphics[scale=0.8]{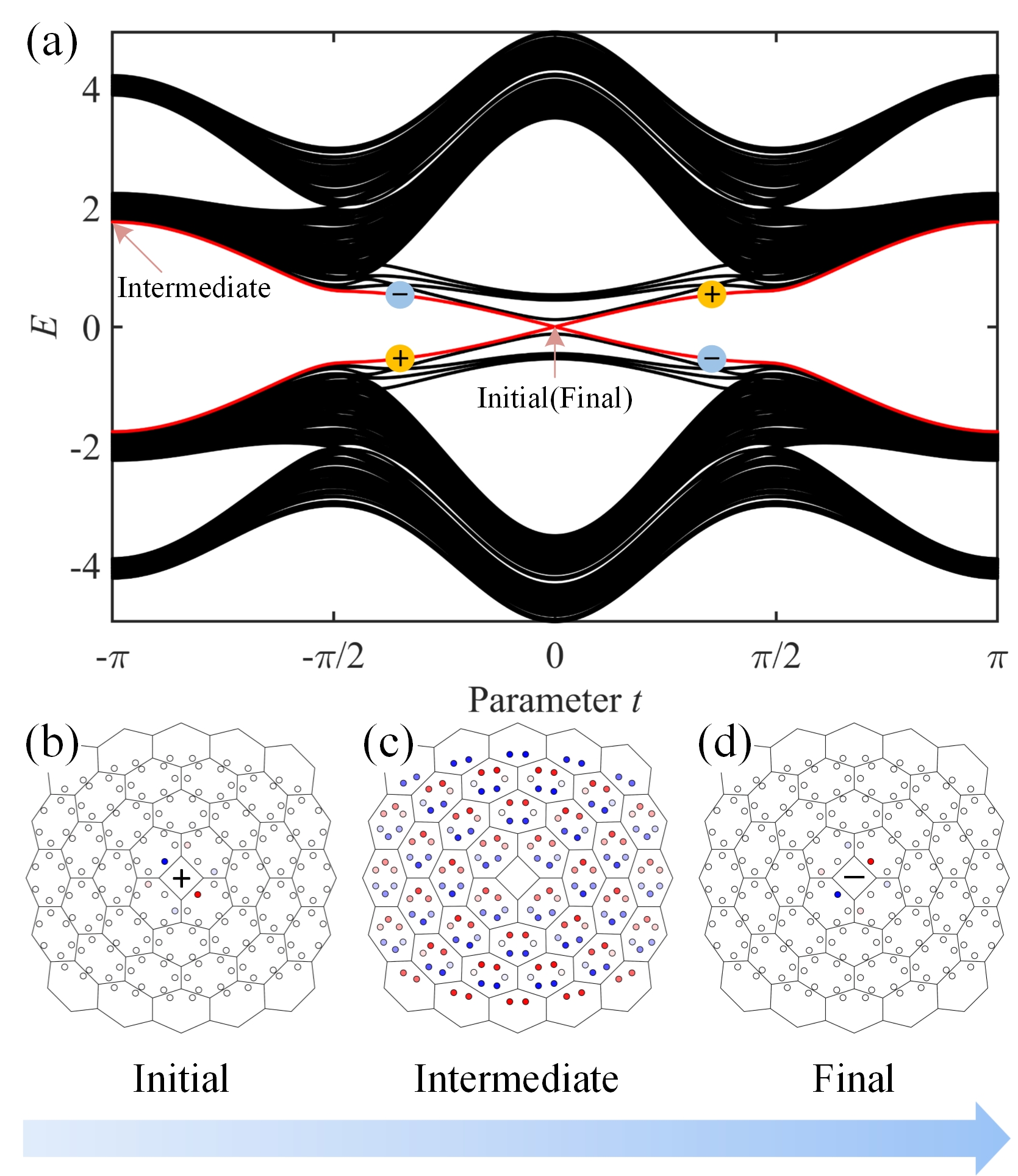}
\captionsetup{format=plain, justification=raggedright}
\caption{Spectral flow of the local charge pumping. (a) The spectral flow of the topological pump of a single disclination core. The disclination states (red line) with different chiralities (labeled by $+$ and $-$) intersect to each other in the pumping process. (b) Local field distribution of the initial state with a positive chirality in Fig. 1(d) . (c) Local field distribution of the intermediate state at the top of the band gap. (d) Local field distribution of the final states with a negative chirality in Fig. 1(d).}
\label{fig:1}
\end{figure}

{\it{Non-local topological pump between two disclination cores}}.---A crucial difference between the disclination states and topologically trivial defect states at the core is that the former is purely induced by the bulk topology (the bulk-disclination correspondence~\cite{BDC3}) while the latter is determined by local structures. As a consequence, the topological pump between disclination states is not limited to those located in the same disclination core. Instead, it can happen between two disclination cores connected by a single bulk topology. To demonstrate this feature, we design a DS with two disclination cores as shown in Fig. 4(a) which have the sublattice symmetry (See details of the construction in Section III in SI~\cite{SUP}). This DS with two disclination cores has the same parameters as Fig. 1 and Fig. 3. Hence it shares the same non-trivial bulk topology as before. Following a similar procedure, we first neglect the onsite energies and find there are 8 disclination states strongly localized at the cores with 4 states being the eigenstates of chiral symmetry.

However, since there is a mirror symmetry between the left sub-deformed structure (sub-DS) and the right sub-DS, four chiral symmetric disclination states forming two sets of chiral partners are degenerate at the zero-energy with field distributions localized at both the left core and the right core (see detailed discussion in Section III in SI~\cite{SUP}). This will lead to the pumping of particles from one core to both of these two cores (see Section III in SI~\cite{SUP}). To avoid this situation, we add a small perturbation to four sites at disclination cores to slightly break the local chiral symmetry at each core. Specifically, we add an extra onsite energy of $0.5\mu$ at four sites (covered by green discs shown in Fig. 4(a)) in the DS where two sites belong to the $A$ sublattice at the left core and two sites belong to the $B$ sublattice at the right core. As a consequence, the above four-fold degenerate disclination states are split into two sets of two-fold degenerate disclination states. In each set, the two degenerate disclination states have opposite chiralities and are localized at different cores (see Fig. 4(b) and detailed discussion in Section IV in SI~\cite{SUP}).

Now we turn on the time modulation in the same way as shown in Fig. 2. We find there is a similar spectral flow during the pumping process. To show topological pump between dislination states at different cores, we firstly prepare an initial zero-energy disclination state with chirality $+1$ localized at sublattice A on the right disclination core (Fig. 4(c)) at the beginning of the pumping process. As $t$ increases, disclination states are gradually pushed into the bulk spectrum (Fig. 4(d)). At the end of the pumping process, disclination states emerge from the bulk spectrum with an opposite chirality $-1$ localized at the $B$ sublattice on the left disclination core (Fig. 4(e)). This non-local topological disclination pump clearly reveals the deep connection between the bulk topology and localized disclination states distinguishing them from trivial defect states which are fully determined by the local geometric structure. 

\begin{figure}
\centering
\includegraphics[scale=0.8]{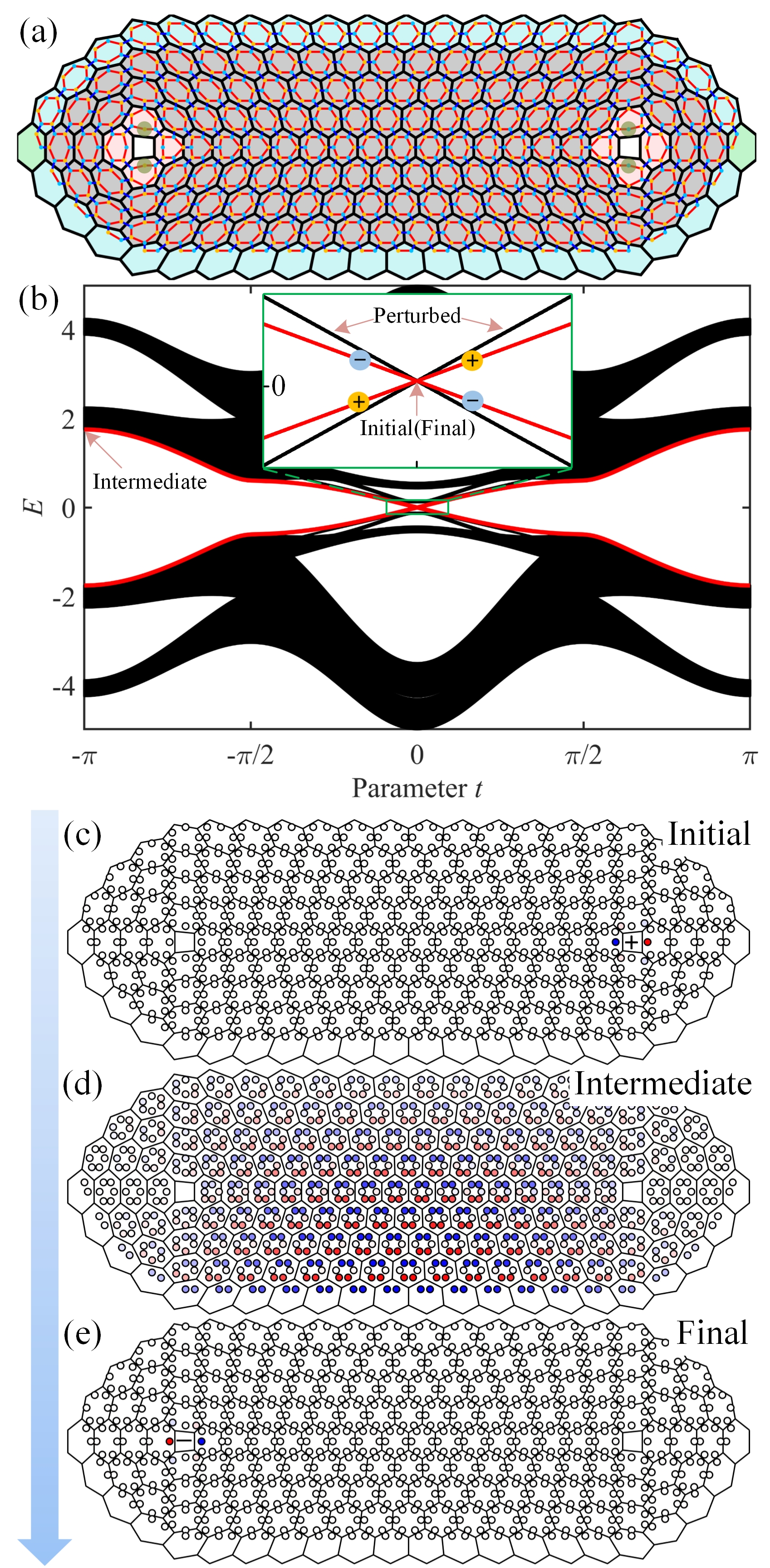}
\captionsetup{format=plain, justification=raggedright}
\caption{The non-local topological pumping. (a) The lattice structure of two dislination cores with same parameters as in Fig. 1(c) with only four sites (sites in green discs) perturbed by adding an onsite energy $0.5\mu$. (b) Spectral flow of (a) with the same parameters in Fig. 3(a). Due to the perturbation, two sets of four disclination states forming chiral symmetry partners with opposite chiralities (inset). (c) The local field distribution of the initial state which localized at the right disclination core with positive chirality. (d) The local field distribution of the intermediate bulk state. (e) The local field distribution of the final disclination state which localized at the left disclination core with negative chirality.}
\label{fig:1}
\end{figure} 

{\it{Conclusion and disscussion}}.---In summary, we proposed a topological pump between disclination states characterized by the evolution of positions of Wannier centers. Moreover, we demonstrate the topological pump both within a single disclination core and between two disclination cores. Our findings generalize the topological pump from conventional lattice structures to 2D inhomogeneous structures with topological defects which can be realized in many classical wave systems such as cold atoms~\cite{cold}, acoustics~\cite{acoustic} electric circuits~\cite{circuit} and the light propagation in femtosecond laser writing waveguide array~\cite{HTP}. Compared to previous studies on topological pump, our result shows many unique features. For example, in conventional topological pump, edge states and corner states are topologically protected by strict crystalline symmetries and the lattice translational symmetry which are vulnerable to inhomogeneous distortions. However, in our case, as long as Wannier centers exist and their motion follows a non-trivial loop, the topological pump is preserved regardless of distortions. Secondly, all previously studied topological pumps are based on outside boundary states such as corner states and edge states whose positions are restricted by the boundary geometry. However, as demonstrated in our paper, we can arbitrarily design positions of disclination cores in structures and therefore our results provide a more flexible way to achieve robust energy transport.




{\sl Acknowledgments}.
This work was financially supported by the University of Hong Kong.

\balance


\begin{thebibliography}{References}

\bibitem{Thouless} D. J. Thouless, Phys. Rev. B \textbf{27}, 6083-6087 (1983).

\bibitem{AS}C. Rorres, Journal of hydraulic engineering, \textbf{126}, 72-80 (2000).

\bibitem{Niu1}Q. Niu and D. J. Thouless, J. Phys. A \textbf{17}, 2453 (1984).

\bibitem{Niu2}Q. Niu, Phys. Rev. Lett. \textbf{64}, 1812-1815 (1990).

\bibitem{pump4}B. Kaestner, et al. Phys. Rev. B \textbf{77}, 153301 (2008).

\bibitem{pump5}P.W. Brouwer, Phys. Rev. B \textbf{58}, 10135 (1998).

\bibitem{pump1}B. L. Altshuler and L. I. Glazman, Science \textbf{283}, 1864-1865 (1999).

\bibitem{pump2}M. Switkes, C. M. Marcus, K. Campman, and A. C. Gossard, Science \textbf{283}, 1905-1908 (1999). 

\bibitem{pump3}M. D. Blumenthal,  et al. Nat. Phys. \textbf{3}, 343-347 (2007). 

\bibitem{pump6}Y. E. Kraus, Y. Lahini, Z. Ringel, M. Verbin, and O. Zilberberg, Phys. Rev. Lett. \textbf{109}, 106402 (2012).

\bibitem{TP0}L. Wang, M. Troyer, and X. Dai, Phys. Rev. Lett., \textbf{111}, 026802 (2013).

\bibitem{TP1}M. Verbin, O. Zilberberg, Y. Lahini, Y. E. Kraus, and Y. Silberberg, Phys. Rev. B, \textbf{91}, 064201 (2015).

\bibitem{TPE1}M. Lohse, C. Schweizer, O. Zilberberg, M. Aidelsburger,  and I. Bloch, Nat. Phys. 12, 350-354 (2016).

\bibitem{TPE2}S. Nakajima, T. Tomita, S. Taie, T. Ichinose, H. Ozawa, L. Wang, M. Troyer, and Y. Takahashi, Nat. Phys. \textbf{12}, 296-300 (2016).

\bibitem{TPE3}J. Tangpanitanon, V. M. Bastidas, S. Al-Assam, P. Roushan, D. Jaksch, and D. G. Angelakis, Phys. Rev. Lett. \textbf{117}, 213603 (2016).

\bibitem{4D1}O. Zilberberg, S. Huang, J. Guglielmon, M. Wang, K. P. Chen, Y. E. Kraus, and M. C. Rechtsman, Nature \textbf{553}, 59-62 (2018).

\bibitem{4D2}M. Lohse, C. Schweizer, H. M. Price, O. Zilberberg, and I. Bloch, Nature \textbf{553}, 55-58 (2018).

\bibitem{TPE4}M. I. Rosa, R. K. Pal, J. R. Arruda, and M. Ruzzene, Phys. Rev. Lett. \textbf{123}, 034301 (2019).

\bibitem{HTP}W. A. Benalcazar, J. Noh, M. Wang, S. Huang, K. P. Chen,  and M. C. Rechtsman, arXiv preprint arXiv:2006.13242.

\bibitem{TD}W. A. Benalcazar, J. C. Y. Teo, and T. L. Hughes, Phys. Rev. B \textbf{89}, 224503 (2014).
 
\bibitem{acoustic1}W. Cheng, E. Prodan, and C. Prodan, Phys. Rev. Lett. \textbf{125}, 224301 (2020). 

\bibitem{acoustic2}X. Xu, Q. Wu, H. Chen, H. Nassar, Y. Chen, A. Norris, M. R. Haberman, and G. Huang, Phys. Rev. Lett. \textbf{125}, 253901 (2020). 

\bibitem{TI1} M. Z. Hasan, and C. L. Kane, Rev. Mod. Phys. \textbf{82}, 3045 (2010).

\bibitem{TI2}X.-L. Qi, and S.-C. Zhang, Rev. Mod. Phys. \textbf{83}, 1057 (2011).

\bibitem{TI3}L. Fu, Topological crystalline insulators. Phys. Rev. Lett. \textbf{106}, 106802 (2011).

\bibitem{TI4}B. Bradlyn, et al. Nature \textbf{547}, 298–305 (2017).

\bibitem{TI5}T. Zhang, et al. Nature \textbf{566}, 475–479
(2019).

\bibitem{TI6}M. G. Vergniory, et al. Nature \textbf{566}, 480–485 (2019).

\bibitem{TI7}F. Tang, H. C. Po, A. Vishwanath, and X. Wan, Nature \textbf{566}, 486–489 (2019).


\bibitem{BBC1}Y. Hatsugai and T. Fukui, Phys. Rev. B, \textbf{94}, 041102 (2016).

\bibitem{BBC2}F. K. Kunst, E. Edvardsson, J. C. Budich, and E. J. Bergholtz, Phys. Rev. Lett. \textbf{121}, 026808 (2018). 

\bibitem{BBC3}L. Trifunovic, and P. W. Brouwer, Phys. Rev. X, \textbf{9}, 011012 (2019).

\bibitem{BDC1}J. C. Y. Teo, and T. L. Hughes, Phys. Rev. Lett. \textbf{111}, 047006 (2013).

\bibitem{BDC2}T. Li, P. Zhu, W. A. Benalcazar, and T. L. Hughes, Phys. Rev. B \textbf{101}, 115115 (2020).

\bibitem{BDC3}Y. Liu, S. Leung, F. F. Li, Z. K. Lin, X. Tao, Y. Poo,  and J. H. Jiang, Nature, \textbf{589}, 381-385 (2021).

\bibitem{BDC4}C. W. Peterson, T. Li, W. Jiang, T. L. Hughes, and G. Bahl, Nature, \textbf{589}, 376-380 (2021).

\bibitem{topo1}F. Liu, H. Y. Deng, and K. Wakabayashi, Phys. Rev. Lett. \textbf{122}, 086804 (2019).

\bibitem{topo2}W. A. Benalcazar, T. Li, and T. L. Hughes, Phys. Rev. B \textbf{99}, 245151 (2019).

\bibitem{topo3}T. Li, P. Zhu, W. A. Benalcazar, and T. L. Hughes, Phys. Rev. B, \textbf{101}, 115115 (2020).


\bibitem{SUP} See Supplementary Information for discussions
about topological phase transition characterized by positions of Wannier centers; Field distributions of corner, edge, and bulk states;	Lattice structure, eigenvalues, and eigenstates of two unchanged disclination cores; Eigenvalues of two disclination cores with removing of sites and perturbation.

\bibitem{cold}S. Chu, Nature \textbf{416}, 206-210 (2002).

\bibitem{acoustic}Z. Yang, F. Gao, X. Shi, X. Lin, Z. Gao, Y. Chong,  and B. Zhang, Phys. Rev. Lett., \textbf{114}, 114301 (2015).

\bibitem{circuit}S. Liu, et al. Light. Sci. \& Appl. \textbf{9}, 1–9 (2020).






\end{thebibliography}
\end{document}